\documentclass[journal=aanmf6,manuscript=article]{achemso}
\pdfoutput=1

\usepackage{achemso}
\usepackage{url}

\usepackage{tabularx}
\usepackage{graphicx}
\usepackage{subfig}
\usepackage{multirow}
\usepackage{makecell}
\usepackage{upquote}

\newcommand*{\qt}{\textquotesingle \textquotesingle}

\newcolumntype{Y}{>{\raggedleft\arraybackslash}X}
\newcolumntype{Z}{>{\centering\arraybackslash}X}

\title[]{Phase Stability and Raman/IR Signatures of Ni-Doped MoS$_2$ from Density-Functional Theory Studies}

\author{Enrique Guerrero} 
\affiliation{Department of Physics, University of California, Merced, Merced, CA 95343}

\author{Rijan Karkee}
\affiliation{Department of Physics, University of California, Merced, Merced, CA 95343}

\author{David A. Strubbe}
\email{dstrubbe@ucmerced.edu}
\affiliation{Department of Physics, University of California, Merced, Merced, CA 95343}

\begin{document}

\begin{abstract}
Ni-doped MoS$_2$ is a layered material with useful tribological, optoelectronic, and catalytic properties. Experiment and theory on doped MoS$_2$ has focused mostly on monolayers or finite particles: theoretical studies of bulk Ni-doped MoS$_2$ are lacking and the mechanisms by which Ni alters bulk properties are largely unsettled. We use density functional theory calculations to determine the structure, mechanical properties, electronic properties, and formation energies of bulk Ni-doped 2H-MoS$_2$ as a function of doping concentration. We find four meta-stable structures: Mo or S substitution, and tetrahedral (t-) or octahedral (o-) intercalation. We compute phase diagrams as a function of chemical potential to guide experimental synthesis. A convex hull analysis shows that t-intercalation (favored over o-intercalation) is quite stable against phase segregation and in comparison with other compounds containing Ni, Mo, and S; the doping formation energy is around 0.1 meV/atom. Intercalation forms strong interlayer covalent bonds and does not increase the $c$-parameter. Ni-doping creates new states in the electronic density of states in MoS$_2$ and shifts the Fermi level, which are of interest for tuning the electronic and optical properties. We calculate the infrared and Raman spectra and find new peaks and shifts in existing peaks that are unique to each dopant site, and therefore may be used to identify the site experimentally, which has been a challenge to do conclusively.
\end{abstract}

\date{\today}
\maketitle

Transition metal dichalcogenides (TMDCs) are lamellar materials with strong covalent intralayer bonds and weak Van der Waals interlayer bonds. MoS$_2$ is a semiconducting TMDC with interesting optical,\cite{Bernardi, Molina} electronic,\cite{Bernardi} spintronic,\cite{Sanikop, Latzke} lubrication,\cite{Vazirisereshk} and catalysis\cite{Mao, Wang, Mosconi} properties, which are often controlled using dopants. Doping MoS$_2$, especially with Ni, can increase catalytic activity and reduce friction.\cite{Stupp, Vellore} However, the mechanisms for these doping effects, and even the basic question of the sites occupied by Ni atoms, remain unclear. This work focuses on exploring Ni-doping in bulk 2H-MoS$_2$ and the effects on materials properties as computed by density functional theory (DFT).

Much previous work about MoS$_2$, especially doping, has been motivated by catalysis.
MoS$_2$ has properties that are desirable in photo-, electro-, and thermocatalysis.\cite{Mao}
Though MoS$_2$ has poor intrinsic catalytic activity, it can be enhanced by dopants\cite{Tedstone} to create defects (especially at edge sites\cite{Sun, Rangarajan, Lauritsen, Wambeke}) as active sites.
Co- and Ni-doping have been studied theoretically\cite{Sun, Hakala} and experimentally\cite{Kondekar, Mosconi, Lauritsen} showing enhancement of the hydrogen evolution reaction.
Intercalation by Na, Co, Ni, and Ca has also been shown to increase catalytic activity in 1T-MoS$_2$.\cite{Attanayake}

More recently, monolayer MoS$_2$ has shown promise for optoelectronic applications such as photovoltaics\cite{Tsai} or LEDs.\cite{LiPan}
Large exciton binding energies and long lifetimes of excitons in MoS$_2$ mean that the high absorption rate can be used to generate useful excitons at room temperature.\cite{Tsai, Bernardi2, Zhong, Schaibley}
Dopants can alter the Fermi energy by donating or accepting electrons and can tune these optoelectronic properties.\cite{Bernardi}
Monolayer and bulk 3R MoS$_2$ lack inversion symmetry, unlike the bulk 2H phase, giving rise to spin-orbit effects which can further be exploited for spintronic and valleytronic applications.\cite{Sanikop, Schaibley}
Transition metal doping by Mn, Fe, Co, and Zn has been used to induce magnetism which could be used for spintronics.\cite{Cheng, Tedstone}

Bulk MoS$_2$ also has useful properties for the older application of lubrication: superior resistance to wear (the gradual loss of material caused by sliding\cite{Vellore}) and a low coefficient of friction (high lubricity) due to the ease of shearing along the basal plane.\cite{Martin}
As a solid lubricant, MoS$_2$ holds some advantage over liquid lubricants---notably it lubricates at temperatures and pressures low enough even for space applications.\cite{Vazirisereshk}
Tribological properties can be enhanced via doping.\cite{Tedstone} Re (in MoS$_2$ fullerene-like nanoparticles\cite{Rapoport}) and Ta\cite{Stupp} have been observed to increase lubricity, and Cr and Ti have been found to strengthen the resistance to humidity while retaining MoS$_2$'s high lubricity.\cite{Ding}
Ni is an exceptional dopant for increased wear protection, decreased coefficient of friction, and long-term stability.\cite{Stupp}
Despite how long studied these materials are, correlation of macroscopic tribological properties and microscopic atomic structure has remained elusive, especially for 2D materials.\cite{Hu, Filleter, Vazirisereshk}

Previous studies suggest that Ni substitutes Mo at edge sites in small flakes,\cite{Sun, Rangarajan, Lauritsen, Wambeke} but there has been little investigation of Ni-doping in bulk.
Some possibilities are suggested by studies of other dopants.
Co\cite{Park} (chemically similar to Ni) and Ta\cite{Zhu} are thought to increase catalytic activity by insertion into S vacancies near the usually inert basal plane.
In bulk, other ions\cite{Julien, Zou, Zhang} can intercalate between the layers or even form ordered alloys.\cite{ChenIntercal}
A phase change from 2H to 1T with Li intercalation has been found via Raman spectra indicating a symmetry change from $D_{6h}$ to $D_{3d}$ and increased interlayer spacing.\cite{Julien}
Ni-doping, presumably through surface energies, can control MoS$_2$ crystal size and growth rates.\cite{Kondekar, Mosconi} A study of Ni-doped MoS$_2$ nanostructures found signs of Ni substitution for Mo and a change to the 3R phase; shifts in diffraction peaks indicated contraction of the MoS$_2$ cell.\cite{Mosconi}

Vibrational spectroscopy has been a main tool for the characterization of 2D materials and their defects.
$n$-type doping of MoS$_2$ is correlated with redshifts of the $A_{\rm 1g}$ Raman-active peak and $p$-type with blueshifts while the $E_{\rm 2g}^1$ Raman frequency remains unchanged.
This behavior is found in field-effect doping as well.\cite{Chakraborty}
With other transition metals, Nb-doping was found to cause an $A_{\rm 1g}$ redshift\cite{Suh} while Au caused a blueshift.\cite{Shi} An $A_{\rm 1g}$ redshift around 1 cm$^{-1}$ for Ni-doped MoS$_2$ nanostructures was attributed to changes in number of layers.\cite{Mosconi}
A new 280 cm$^{-1}$ peak was found in Ni-doped MoS$_2$ that was not related to either pure MoS$_2$ or nickel sulfides,\cite{Kondekar} but its origins have not been further characterized.
Vibrational properties are also important for tribology as energy dissipation via vibrational states\cite{Tangney, Hu} and electron-phonon coupling\cite{Filleter} (EPC) are thought to play a crucial role in increasing friction in 2D materials.
We have not found previous vibrational spectra computations of doped bulk 2H-MoS$_2$ and experimental reports have rarely had sufficient resolution to identify dopant-related peaks.
There is a clear need for more detailed Raman and IR studies since vibrational spectroscopy can be a powerful tool for characterizing defects.

In this paper, we use first principles DFT calculations to systematically survey different structures and concentrations of bulk Ni-doped MoS$_2$ to definitively resolve the structure of this material.
We determine phase stability to guide synthesis and show unusual stability of intercalated structures.
We compute Raman and IR spectra to aid in experimental identification of defect structures (which may not be lowest energetically due to non-equilibrium growth), and study elastic and vibrational properties to understand how Ni improves performance in lubrication.
By studying elasticity, bonding, and vibrations, we elucidate how Ni decreases friction and increases resistance to wear.

\section{Results and Discussion}
\begin{figure}
	\includegraphics[width=400px]{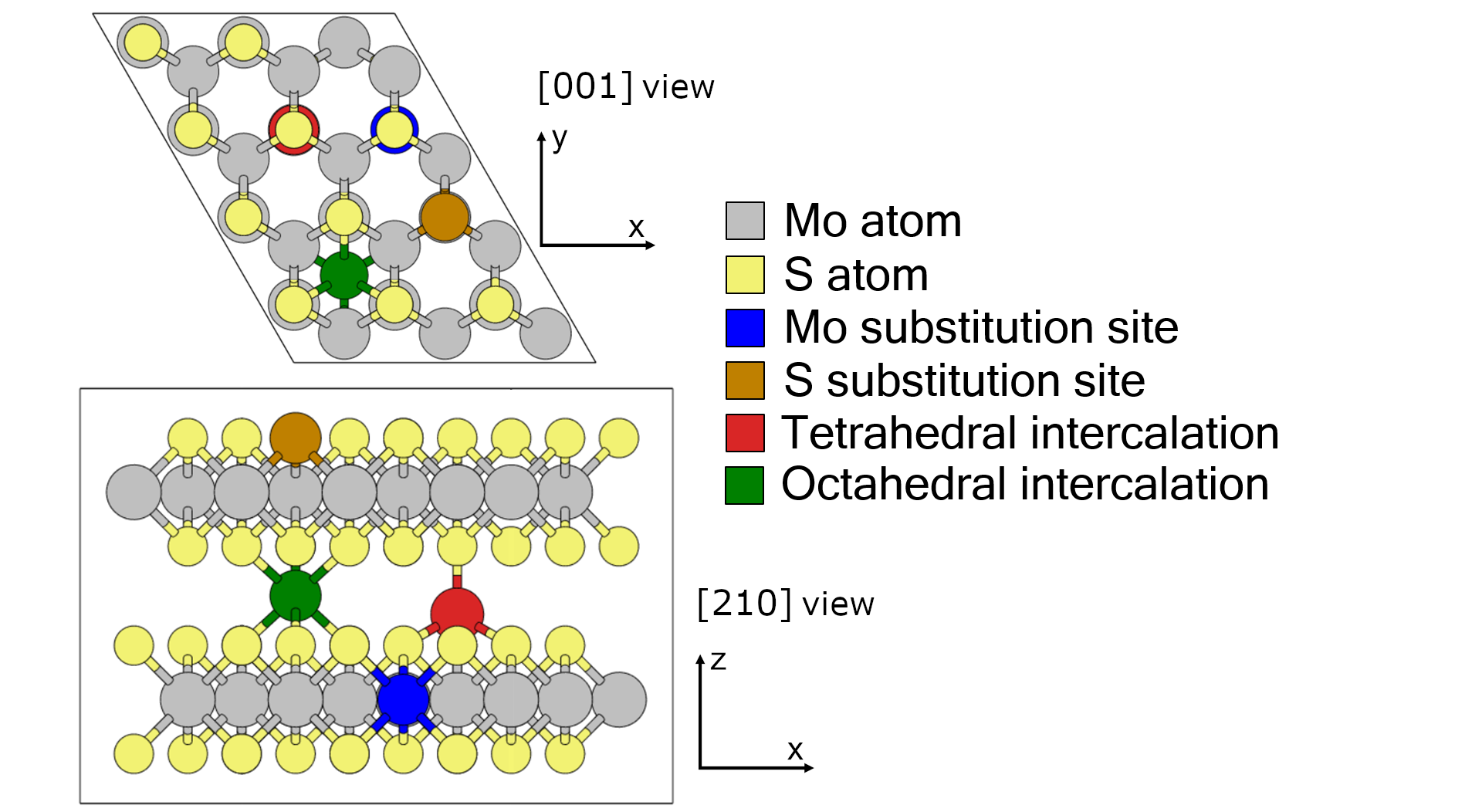}
	\caption{Possible sites for Ni dopants in 2H-MoS$_2$: Mo or S substitution, or intercalation at tetrahedral (t) and octahedral (o) interlayer sites. Other highly symmetric sites (bridge and intralayer interstitials) were found to be unstable.}
	\label{fig:Dopants}
\end{figure}

\begin{table*}[t]
	\begin{tabularx}{460px}{ | c | c | c | Z | c | c | Z | }
		\hline
		&
		\multicolumn{3}{c | }{This Work} &
		\multicolumn{3}{c | }{Literature} \\
		\hline &
		{\bfseries LDA } &
		{\bfseries PBE } &
		{\bfseries PBE + GD2} &
		{\bfseries Exp't} &
		{\bfseries LDA} &
		{\bfseries PBE + vdW}$^\dagger$ \\
		\hline
		\hline
		$a$ (\AA) &	3.12 & 3.18 & 3.19 & 3.16\cite{Wilson} & 3.13\cite{Molina2} & 3.16\cite{Yengejeh}\\
		{$c$/$a$} &	3.87 & 4.64 & 3.89 & 3.89\cite{Wilson} & 3.89\cite{Molina2} & 3.89\cite{Yengejeh}\\
		$E_{\rm 2g}^1$ (cm$^{-1}$) & 389 & 373 & 371 & 383\cite{Livneh} & 388\cite{Molina2} & 372\cite{Coutinho}\\
		$A_{\rm 1g}$ (cm$^{-1}$) & 413 & 402 & 403 & 409\cite{Livneh} & 412\cite{Molina2} & 397\cite{Coutinho} \\
		$E_{\rm 1u}$ (cm$^{-1}$) & 390 & 373 & 372 & 384\cite{Livneh} & 391\cite{Molina2} & 372\cite{Coutinho} \\
		$A_{\rm 2u}$ (cm$^{-1}$) & 471 & 458 & 457 & 470\cite{Livneh} & 469\cite{Molina2} & 454\cite{Coutinho} \\
		$C_{11}$ (GPa) & 242 & 181 & 212 & 238\cite{Feldman} & 240\cite{Todorova} & 223\cite{Yengejeh} \\
		$C_{33}$ (GPa) & 53 & 1.70* & 51 & 52\cite{Feldman} & 53\cite{Todorova} & 49\cite{Yengejeh} \\
		C$_{55}$ (GPa) & 20 & 0.76* & 15 & 19\cite{Feldman} & 32\cite{Todorova} & 15\cite{Yengejeh} \\
		\hline
	\end{tabularx}
	\caption{Comparison of Lattice Parameters, Vibrational Frequencies at $q=0$, and Elastic Parameters}
	*PBE without Van der Waals underestimates the elasticity in the $z$-direction.

	$^\dagger$Ref.\cite{Coutinho} seems to use Tkatchenko-Scheffler\cite{TS} dispersion corrections while Ref. \cite{Yengejeh} uses Grimme-D3.\cite{GD3}
	\label{table:Fnl}
\end{table*}

We consider a set of possible dopant sites shown in Fig. 1: Mo substitution, S substitution, octahedral (o-) intercalation, and tetrahedral (t-) intercalation, as studied for Nb-doped MoS$_2$.\cite{Ivanovskaya} Other possible sites included the bridge site above an Mo-S bond and in between two layers, and the hollow-site interstitial directly between three S atoms on the S plane. These were ruled out as possibilities because they were found to be unstable under relaxations---they both relaxed to o-intercalation.

We consider doping of the most common MoS$_2$ structure: 2H, belonging to the D$_{6h}$ point group with 6 atoms per unit cell, which is very close in energy to the 3R bulk phase.\cite{ChenPhaseTrans} Introducing Ni lowers symmetry: Mo substitution, S substitution, o-, and t-intercalation structures belong to the $D_{3h}$, $C_{3v}$, $D_{3d}$, and $C_{3v}$ point groups respectively. We considered doped structures of increasing MoS$_2$ supercell size in-plane from 1$\times$1 to 4$\times$4, each with 1 Ni atom per cell, with Ni concentrations of 16.7 at\%, 4.2 at\%, 1.9 at\%, and 1.0 at\%. Decreasing Ni concentration allows us to extrapolate quantities to the low-doping limit.

Van der Waals interactions are important between MoS$_2$ layers, but they are a challenge for DFT calculations. Since such interactions are almost absent in PBE, we considered PBE with and without Grimme-D2 (GD2)\cite{GD2} Van der Waals corrections. We found Grimme-D3\cite{GD3} to give very similar structures to GD2. However, GD2 and related approaches are particularly poor for metals.\cite{Andersson} This is not an issue for MoS$_2$, but it is problematic for computing formation energies with reference to pure Ni and Mo. By contrast, LDA is quite accurate for metals, and also successful for Van der Waals-bonded quasi-2D systems: empirically, it is seen that LDA's overestimation of interatomic interactions balances the lack of explicit non-local Van der Waals interactions.\cite{Yin} Therefore we focus on LDA for formation energies such as in Table \ref{table:EnAboveHull}.
We also find that LDA provides better agreement with experimental vibrational frequencies than PBE or PBE+GD2 (Table 1).

With these considerations, we have chosen to use LDA calculations for energies in the phase diagrams, convex hull, and vibrational calculations. As shown by Peelaers and Van de Walle, adding dispersion corrections to PBE can give reasonable structures and elastic properties;\cite{Peelaers, PeelaersVdW} we confirm this in Table \ref{table:Fnl}. PBE+GD2 was thus used for structural parameters, elastic parameters, and electronic densities. PBE with no correction was found to be quantitatively worse than LDA and is not used except as a point of comparison: the energy landscape is very flat with respect to changes in layer separation, leading to large errors in the lattice parameter $c$ and the $C_{33}$ and $C_{55}$ parameters.

We calculated formation energy for pristine and doped MoS$_2$, as well as other crystals containing Mo, S, and/or Ni. We used the Materials Project\cite{MaterialsProject} to find materials with relatively stable computed energies (less than 0.1 eV/atom above hull): Ni$_2$Mo, Ni$_3$Mo, Ni$_4$Mo, NiS$_2$, Ni$_3$S$_4$, NiS, Ni$_9$S$_8$, Ni$_3$S$_2$, MoS$_2$, and Mo$_3$S$_4$, and the Chevrel phase NiMo$_3$S$_4$. Details on the calculation parameters for the structures used can be found in Table S1. NiMo$_3$S$_4$\cite{Guillevic} appears to be the only known stochiometric crystal that contains all three species. It could be a competing material during the synthesis of Ni-doped MoS$_2$, and has been studied for catalysis of the hydrogen evolution reaction.\cite{Wang}

\begin{figure}
	\includegraphics[width=460px]{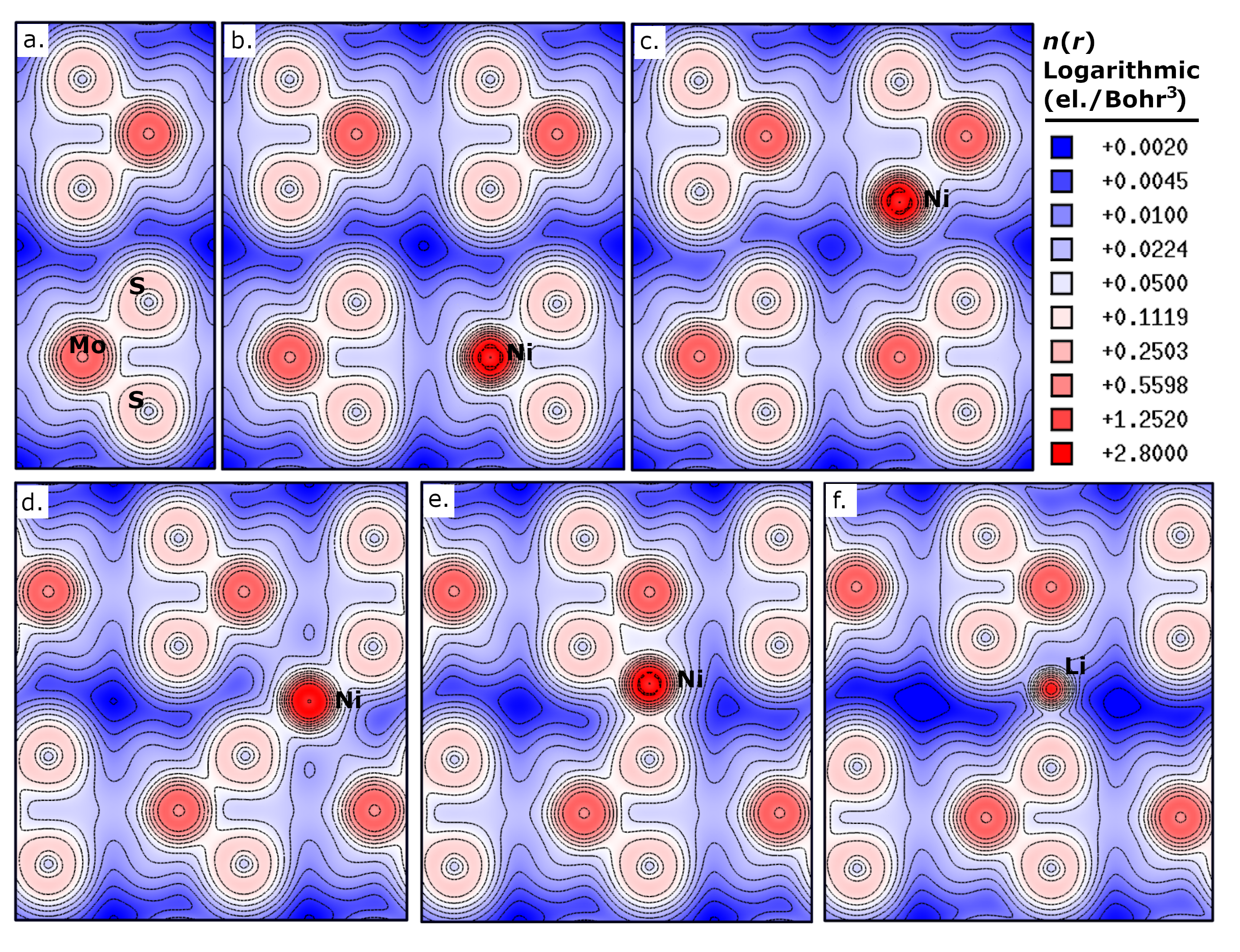}
	\caption{Cross sections of the electron densities of a) pristine MoS$_2$, b) Ni substituting Mo, c) Ni substituting S, d) Ni o-intercalation, e) Ni t-intercalation, and f) Li t-intercalation within 2$\times$2 supercells. The data is displayed in a logarithmic color scale in the $xz$-plane which includes Mo-S and Ni bonds. In intercalated structures, the electron density has strongly increased in the interatomic region between layers. This suggests formation of covalent bonds for intercalated Ni, unlike in the well-known case of Li.\cite{Enyashin} Densities were plotted using XCrysDen.\cite{XCrySDen}} \label{fig:Densities}
\end{figure}

The electron density can be a useful tool to study bonding. The pristine unit cell and doped 2$\times$2 supercell electron densities are shown in Fig. \ref{fig:Densities}. The maximum electron density in the Ni-S or Ni-Mo bonding regions is similar to that of the pristine Mo-S bonds, signifying the Ni bonds are strong and covalent in all cases. The maxima are slightly less than for Mo-S bonds in each case, except they are higher for the Ni-S interlayer bonds of t-intercalation. The surprising result that Ni can readily substitute S with the bonding network intact indicates that Ni, like Co,\cite{Park} could potentially be used to fill S vacancies in a stable structure. Results from formation energies with respect to vacancies, however, show that filling an S vacancy with Ni requires energy while filling a Mo vacancy with Ni releases energy (Table S6).

The o- and t-intercalated structures show clear layer-to-layer bonds with the Ni atom as a bridge. Li is a known MoS$_2$ intercalant that widens the separation between layers\cite{Enyashin} and is stripped of its outer electron, leaving a closed-shell ion that does not form covalent bonds, as shown in Fig. \ref{fig:Densities}f. Comparing the Li case to Ni shows a stark contrast in the electron density, indicating that Ni forms strong covalent interlayer bonds as opposed to Van der Waals or ionic interactions for Li. The interlayer density is unchanged from the pristine in the Mo-substituted case but has increased significantly in the case of S substitution, indicating increased interlayer interactions.

We find a small reduction in the $c$-parameter for Mo substitution, as well as for S substitution and o-intercalation, especially at high concentration. This is consistent with Mosconi \textit{et al.}'s report that Ni-doping results in reduction of the $c$-parameter, attributed to Mo substitution and Ni's smaller covalent radius.\cite{Mosconi} The basal lattice parameter $a$/$b$ is essentially unchanged, however, regardless of concentration. 3$\times$3 and 4$\times$4 supercells of Mo-substituted and o-intercalation structures show a pseudo-Jahn-Teller symmetry-breaking in their Ni-S bond lengths. When Ni substitutes Mo, four Ni-S bonds are 2.2 \AA\ long (2 bonds to each of the S planes), an additional bond is slightly longer by 0.2 \AA\ and the final S atom has been pushed away significantly to a distance of 3.3 \AA . A similar symmetry-breaking has been calculated for Mo substitution on the basal plane.\cite{Hakala} A fully symmetric o-intercalation structure is possible, unlike the t-intercalation which automatically breaks symmetry between layers. Despite this, o-intercalation breaks symmetry: Ni-S bonding distances in the closer layer are 0.2 \AA\ shorter than those in the further layer. A full concentration dependent set of atomic distances and cell parameters can be found in Table S2.

\begin{figure}
	\includegraphics[width=200px]{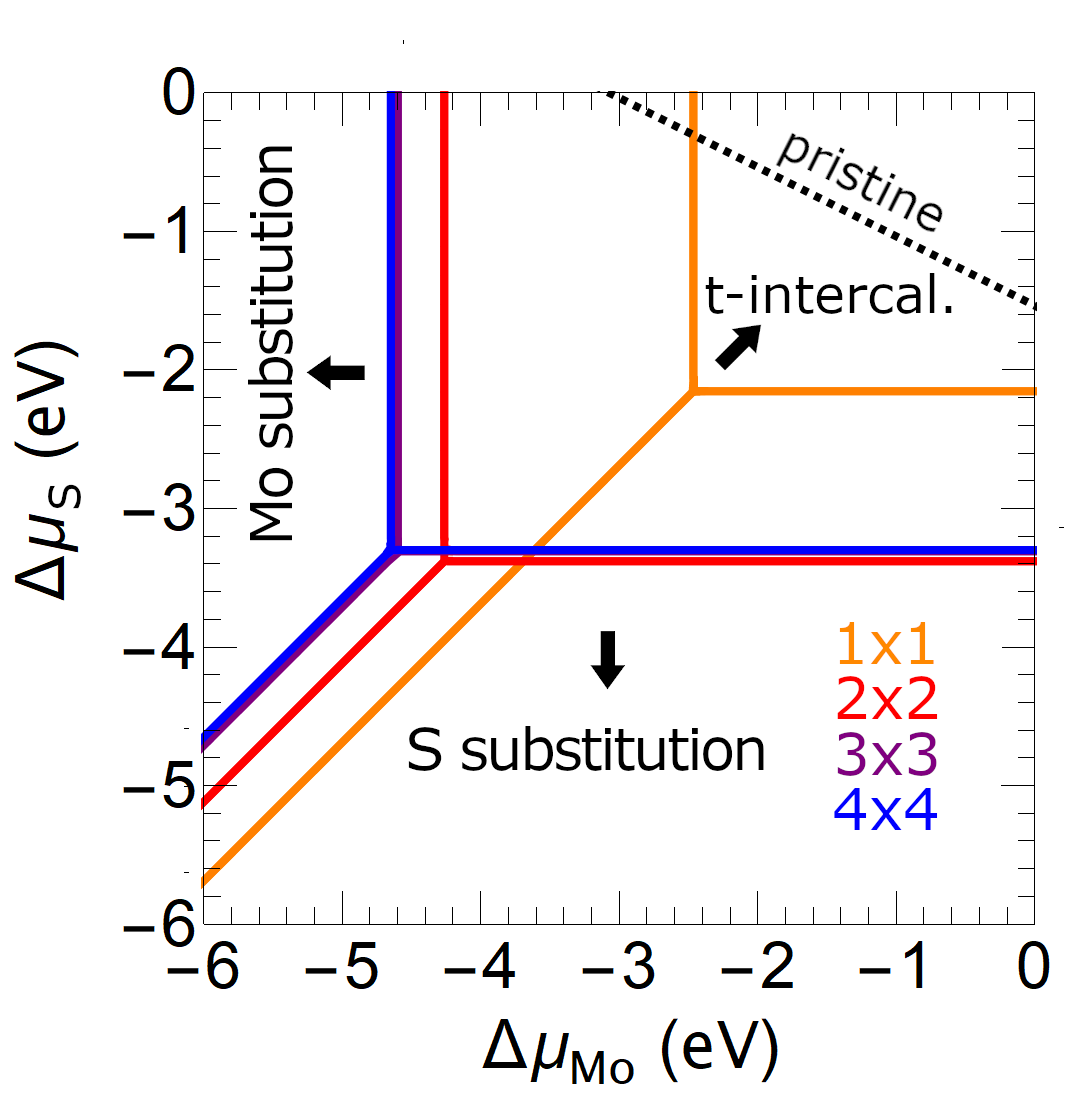}
	\caption{Thermodynamically favored doped structures at different chemical potentials relative to bulk phases, as calculated by LDA. The equilibrium lines change depending on the concentration, but do not change much between LDA, PBE, and PBE+GD2. The dashed line indicates the boundary above which MoS$_2$ is stable. Phase diagrams for PBE and PBE+GD2 can be found in Fig. S1.} \label{fig:chemPot}
\end{figure}
\begin{figure}
	\includegraphics[width=300px]{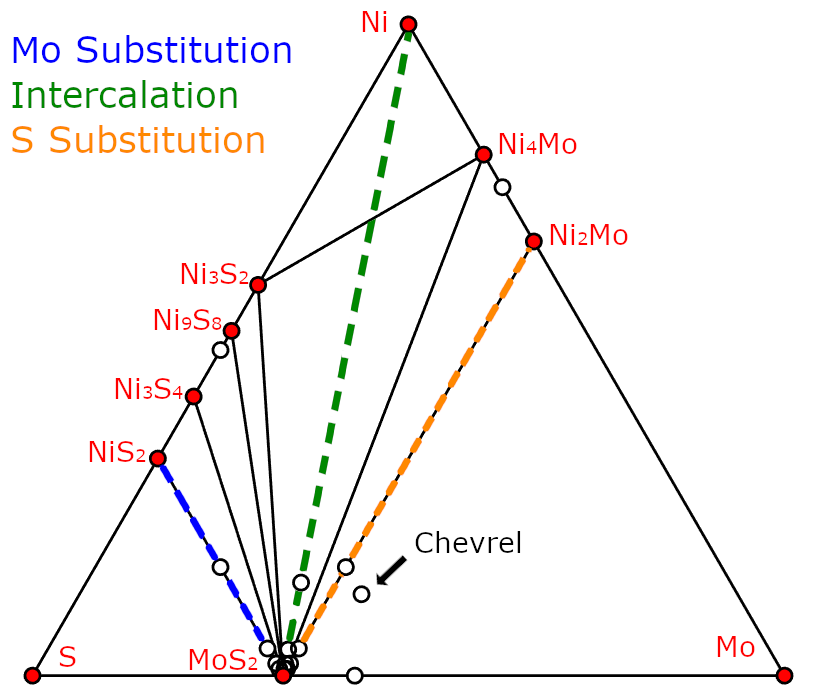}
	\caption{Phase diagram for materials containing Ni, Mo, and S, based on LDA energies. Materials' position along the plane edges indicate their stoichiometry. Materials on the convex hull are shown in red. Structures with energies above  hull are represented as open circles; all the doped structures fall within this category and are mostly clustered around the MoS$_2$ point. The t-intercalation is above but very close to the convex hull.} \label{fig:ConvexHull}
\end{figure}
\begin{table*}[t]
	\begin{tabular} { | c | c || c | c || c | c || c | c | }
		\hline
		Mo & 0 &
		Ni$_3$Mo & 0.024 &
		NiS & 0.018
		& MoS$_2$ & 0 \\
		Ni & 0 &
		Ni$_4$Mo & 0 &
		Ni$_3$S$_4$ & 0 &
		Mo$_3$S$_4$ & 0.065 \\
		S & 0 &
		NiS$_2$ & 0 &
		Ni$_9$S$_8$ & 0 & & \\
		Ni$_2$Mo & 0 &
		NiMo$_3$S$_4$ & 0.050 &
		Ni$_3$S$_2$ & 0 & & \\
		\hline
	\end{tabular}
	\begin{tabular} { | r | c | c | c | c | }
		\hline & Mo subst. & S subst. & t-intercal. & o-intercal. \\
		\hline
		1$\times$1 & 0.149 & 0.173 & 0.025 & 0.116 \\
		2$\times$2 & 0.107 & 0.090 & 0.002 & 0.044 \\
		3$\times$3 & 0.054 & 0.039 & 0.001 & 0.021 \\
		4$\times$4 & 0.031 & 0.022 & 0.001 & 0.012 \\
		\hline
	\end{tabular}
	\caption{Energies per Atom Above Hull in eV, According to LDA, for Computed Solids Containing Mo, S, and/or Ni.} \label{table:EnAboveHull}
\end{table*}

Since the different doped structures have different stoichiometries, we compare their thermodynamic favorability using not only formation energies but also the chemical potentials of the elements with respect to their bulk phases. The favored structures as a function of chemical potentials are shown in Fig. \ref{fig:chemPot}. We focus on the region with $\Delta \mu_{\rm S} \le 0$ and $\Delta \mu_{\rm Mo} \le 0$, where the bulk Mo and S would not be formed, as in stability triangle analysis.\cite{ZhangStabTri} With different supercell sizes, and thus concentrations, the favorable chemical potential regions shift slightly. Except for the very highly-doped 1$\times$1 structure, the triple point rests around $\Delta \mu_{\rm Mo} = -4.0$ eV and $\Delta \mu_{\rm S} = -3.7$ eV. These values can be used to select the relative concentrations of reactants used in the synthesis process to achieve a particular target structure. The proximity of the 3$\times$3 and 4$\times$4 lines show that the 3$\times$3 supercell is converged to the low-doping limit. O-intercalation structures do not appear on this plot because they have the same stoichiometry as t-intercalation, but always have a larger formation energy by about 0.07 eV per Ni atom. For comparison, calculations indicate o-intercalation is preferred over t-intercalation for Nb,\cite{Ivanovskaya} Li,\cite{Enyashin} and Mo.\cite{ZhaoInterc} The stability line\cite{ZhangStabTri} for pristine MoS$_2$ is also shown in Fig. \ref{fig:chemPot}, below which any kind of MoS$_2$ structure is not expected. Since only the t-intercalation region (for supercells greater than 1$\times$1) exists above this line, we conclude that equilibrium growth will yield t-intercalation, and the substituted sites can only be reached out of equilibrium. Phase diagrams for PBE and PBE+GD2 are shown in Fig. S1 and show a similar trend to LDA with respect to increasing supercell sizes. Formation energies \textit{vs.} bulk Mo and S (Table S4) are -0.6 to -1.0 eV, but to judge favorability of doping, we must compare to MoS$_2$. Previous studies\cite{Dolui} have focused on the end-points of this pristine stability line: where $\Delta \mu_{\rm S} = 0$ (``S-rich'') or $\Delta \mu_{\rm Mo} = 0$ (``Mo-rich''). We consider in both cases that $\Delta \mu_{\rm Ni} = 0$, \textit{i.e.} ``Ni-rich'' as well. We find that t-intercalation is the favored structure for both ends at all supercell sizes, except for $1 \times 1$ where S-rich conditions favor Mo-substitution. The positive formation energies \textit{vs.} MoS$_2$ (0.01-0.4 eV) in Table S5 show that doping is unfavorable generally, though t-intercalation is essentially stable with an extremely small formation energy of $\sim$0.1 meV/atom.

The convex hull is used to identify structures that are stable against phase segregation.\cite{Jain, Kutana} Our calculation (Fig. \ref{fig:ConvexHull}) shows that neither the Chevrel nor the doped compounds are the most stable structures at their concentrations. Therefore, these structures will phase-segregate into nickel sulfides, Ni-Mo alloys, and pure elements given enough time and temperature. Table \ref{table:EnAboveHull}, however, shows that the low-concentration doped structures have small energies above hull. The t-intercalation energies above hull, in fact, are below $k_BT$ at room temperature for all concentrations, and have magnitude within the margin of error for DFT energies. Therefore we may regard these structures as stable: synthesizable and unlikely to phase-segregate. Despite being observed at the edges of small flakes,\cite{Lauritsen} Mo substitution has high energies above hull which are generally higher than for S substitution. The Chevrel phase (NiMo$_3$S$_4$) has an energy above hull on the order of the 3$\times$3 Mo- and S-substituted phases---since this phase is known to be experimentally synthesizable,\cite{Guillevic} the results suggest the doped phases will also be accessible. Consideration of 1D convex hull plots along the colored lines of Fig. \ref{fig:ConvexHull} shows very small energy differences above hull for t-intercalation \textit{vs.} MoS$_2$ and bulk Ni, but much larger differences for Mo substitution \textit{vs}. MoS$_2$ and NiS$_2$, or S substitution \textit{vs}. MoS$_2$ and MoNi$_2$ (Fig. S3). Several Ni$_x$S$_y$ phases are seen to be stable, consistent with observation of their forming during synthesis by Kondekar \textit{et al.}\cite{Kondekar} Phase diagrams for PBE and PBE+GD2 and energies above hull are given in Fig. S2 and Table S3.

\begin{table*}[t]
	\begin{tabular} { | c | c | r | r | r | r  | r | }
		\hline (GPa) &  & {\bfseries undoped} & {\bfseries Mo subst.} & {\bfseries S subst.} & {\bfseries t-intercal.} & {\bfseries o-intercal.} \\
		\hline
		\hline
		\multirow{4}{*}{$C_{11}$}
		& 1$\times$1 & 211.8 & 159.2 & 168.9 & 201.4 & 208.2 \\
		& 2$\times$2 & \qt & 177.2 & 204.8 & 211.5 & 210.6 \\
		& 3$\times$3 & \qt & 196.5 & 207.7 & 211.1 & 208.7 \\
		& 4$\times$4 & \qt & 190.8 & 210.0 & 212.0 & 210.2 \\
		\hline
		\multirow{4}{*}{$C_{33}$}
		& 1$\times$1 & 51.3 & 50.0 & 91.7 & 99.4 & 94.1 \\
		& 2$\times$2 & \qt & 51.3 & 57.7 & 76.1 & 72.9 \\
		& 3$\times$3 & \qt & 49.0 & 55.8 & 63.4 & 63.3 \\
		& 4$\times$4 & \qt & 50.0 & 54.3 & 57.9 & 58.7 \\
		\hline
		\multirow{4}{*}{$C_{55}$}
		& 1$\times$1 & 15.1 & 7.5 & 24.3 & 24.0 & 25.3 \\
		& 2$\times$2 & \qt & 13.1 & 10.9 & 19.1 & 20.6 \\
		& 3$\times$3 & \qt & 13.8 & 14.0 & 16.7 & 12.4 \\
		& 4$\times$4 & \qt & 14.2 & 14.6 & 15.9 & 13.2 \\
		\hline
	\end{tabular}
	\caption{Elastic Parameters for Ni-Doped Structures at Different Supercell Sizes.*}
	*1$\times$1 through 4$\times$4 supercells correspond to Ni concentrations of 17 at\%, 4 at\%, 2 at\%, and 1 at\% respectively. Most structures did not change appreciably in $C_{11}$ besides the 20-30 GPa drop for Mo substitution. $C_{33}$ values were appreciably raised by 20 GPa in intercalated cases. $C_{55}$ did not show large changes, giving evidence against the idea that Ni-doping lowers the frictional coefficient by lowering the shear stiffness.
	\label{table:Elastic}
\end{table*}

We found the elastic parameters, important for tribology as the material is under stress, are altered by Ni-doping as shown in Table \ref{table:Elastic}. The in-plane $C_{11}$ parameter is notably weakened in the case of Mo substitution. This may be related to the lower charge density between Ni and S as seen in Fig. \ref{fig:Densities}b. At lower Ni concentrations, the reduced number of bonds leads to a lower $C_{11}$. S substitution also weakened $C_{11}$ slightly, but the intercalated cases saw little change at any supercell sizes beyond 1$\times$1. Out-of-plane $C_{33}$ stiffness increased in the case of intercalated and S-substituted structures due to interlayer bonding with intercalated Ni, and increased interlayer interactions for S substitution. The increased out-of-plane stiffness will contribute to the increased resistance to wear (layers flaking off), as seen in experiments.\cite{Stupp, Vellore}

Since Ni-doping is known to reduce the coefficient of friction of MoS$_2$, we might expect $C_{55}$ would be lowered, reducing the shear stiffness. However, there is not much difference between doped and undoped $C_{55}$ values at low concentrations. Further, at high concentrations (5-15 at\%) $C_{55}$ of intercalated structures is significantly higher than the pristine case (except for Mo substitution, where it is halved). This indicates that a reduction in shear stiffness is not the explanation for the observed reduction in frictional coefficient due to Ni-doping.

\begin{figure}
	\includegraphics[width=460px]{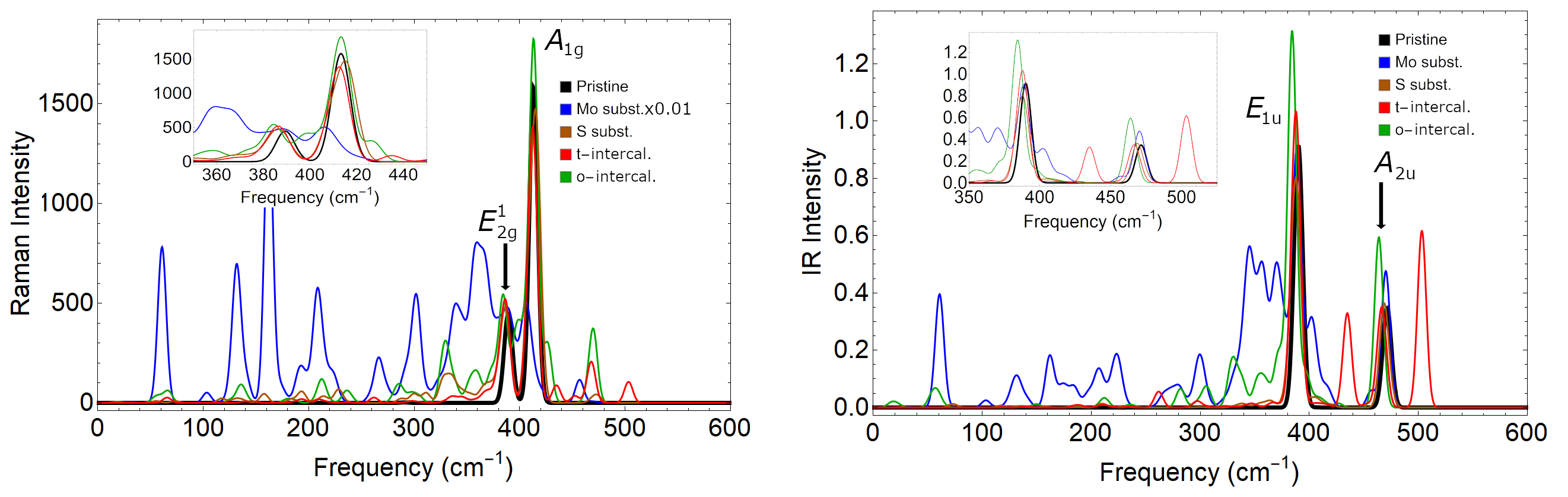}
	\caption{Raman and IR spectra, in A$^4$/amu per MoS$_2$ unit and (D/A)$^2$/amu per MoS$_2$ unit respectively, for 3$\times$3 doped structures. A Gaussian broadening of 4 cm$^{-1}$ is used. Mo-substituted Raman intensities are large, so they are scaled by a factor of 0.01. Low frequency modes ($<$ 100 cm$^{-1}$) correspond to layer breathing and shearing modes. Insets more closely show the pristine Raman-active $A_{\rm 1g}$ and $E_{\rm 2g}^1$ modes and IR active $E_{\rm 1u}$ and $A_{\rm 2u}$ modes. The strong IR and Raman peaks of the pristine MoS$_2$ remain with doping but new peaks appear at different frequencies which can be used for experimental identification of the doping site.}
	\label{fig:VibSpec}
\end{figure}
\begin{table*}[t]
	\begin{tabularx}{470pt}{ | c | c | Y | Y | Y | Y | }
		\hline \multicolumn{2}{|c|}{\bfseries Frequency (cm$^{-1}$)} & {\bfseries Mo subst.} & {\bfseries S subst.} & {\bfseries t-intercal.} & {\bfseries o-intercal.} \\
		\hline
		\hline
		\multirow{9}{*}{Raman}
		& \multirow{4}{*}{New peak} &
		61, 131, 132, 162, 266, 301, 364 &
		327, 333, 337 &
		366, 372, 435, 452, 503  &
		212, 235, 330, 361, 427 \\
		\cline{2-6}
		& \multirow{3}{*}{Activation} &
		209, 457 &
		473 &
		336, 465, 468 &
		67, 135, 287, 287, 470 \\
		\cline{2-6}
		& {$E_{\rm 2g}^1$ shift} &
		-3.2 &
		-3.2 &
		-3.1 &
		-5.4\\
		\cline{2-6}
		& {$A_{\rm 1g}$ shift} &
		-7.1 &
		+2.0 &
		-0.2 &
		-0.5 \\
		\hline
		\multirow{7}{*}{IR}
		& \multirow{4}{*}{New peak} &
		61, 131, 132, 162, 301, 345, 356, 370  &
		- &
		262, 435, 503  &
		330 \\
		\cline{2-6}
		& {Activation} &
		209, 224, 388, 389, 470  &
		473 &
		465, 468 &
		54, 59, 282, 282 \\
		\cline{2-6}
		& {$E_{\rm 1u}$ shift} &
		-0.6 &
		-0.8 &
		-1.9 &
		-6.1 \\
		\cline{2-6}
		& {$A_{\rm 2u}$ shift} &
		-14.5 &
		-3.0 &
		-6.2 &
		-7.5 \\
		\hline
	\end{tabularx}
	\caption{Changes to the Pristine Vibrational Spectra for Doped 3$\times$3 Structures.}
	\label{table:QualitativeShifts}
\end{table*}
\begin{table*}[t]
	\begin{tabular} { | c | r | r | r | r | r | }
		\hline
		{\bfseries (cm$^{-1}$)} & {\bfseries undoped} &	{\bfseries Mo subst.} &	{\bfseries S subst.} & {\bfseries t-intercal.} &	{\bfseries o-intercal.} \\
		\hline
		\hline
		\multicolumn{6}{ | l | }{Shearing-like mode frequency (pristine $E_{\rm 2g}^2$)} \\
		\hline
		1$\times$1 & 35 & 38, 43 & 88, 89 & 57, 57 & 48, 49 \\
		2$\times$2 & \qt & 26, 30 & 24, 27 & 42, 43 & 41, 41 \\
		3$\times$3 & \qt & 32, 34 & 22, 25 & 30, 30 & 17, 22 \\
		\hline
		\multicolumn{6}{ | l | }{Layer breathing-like mode frequency (pristine $B_{\rm 2g}^2$)} \\
		\hline
		1$\times$1 & 56 & 88 & 132 & 127 & 108 \\
		2$\times$2 & \qt & 63 & 58 & 78 & 71 \\
		3$\times$3 & \qt & 59 & 55 & 66 & 67 \\
		\hline
	\end{tabular}
	\caption{Calculated $q=0$ Low-Frequency Vibrational Modes in Pristine and Doped MoS$_2$.}
	\label{table:SlidingShearingFreqs}
\end{table*}

To aid in experimentally identifying Ni-doped MoS$_2$, we have computed the Raman and IR intensities for MoS$_2$ with the different doped sites. Previous experimental work\cite{Mignuzzi, Suh, Mosconi} on doped MoS$_2$ has focused on shifts in the $A_{\rm 1g}$ and $E_{\rm 2g}^1$ Raman peaks relative to the pristine positions,\cite{Iqbal} but has not investigated the formation of new peaks due to the presence of a new atom of a different mass, or symmetry-breaking due to the dopant, as seen in other doped 2D materials;\cite{Yuan, ZhangWS2} one exception is the study of Li-intercalated MoS$_2$ by Sekine \textit{et al.}\cite{Sekine}
Raman and IR spectra are compared in Fig. \ref{fig:VibSpec} and key distinguishing features are listed in Table \ref{table:QualitativeShifts}. Mode characters are detailed in Tables S7-10. We have computed the spectra for 1$\times$1, 2$\times$2, and 3$\times$3 supercells to see the concentration-dependence of the strong Raman and IR active peaks and find that 3$\times$3 is converging to a low-doping limit (Figs. S5 and S6), though significant Raman intensity variations with concentration remain for Mo substitution and o-intercalation.

As in previous studies,\cite{Kondekar, Suh, Vandalon} the doped spectra show peaks corresponding to the pristine ones, with changes in frequency. Shifts in the pristine Raman/IR peaks can be caused by a number of factors: the mass difference of Ni compared to substituted Mo or S, an alteration of bond strengths by $n$- or $p$-type doping, or induced strain.\cite{KimRaman} Ni has two more electrons than Mo and results in $n$-type doping (as discussed below), which would generally be expected to result in a redshift of the $A_{\rm 1g}$ peak.\cite{Iqbal} In Li-intercalated MoS$_2$, Sekine \textit{et al.}\cite{Sekine} experimentally found redshifts of the $E_{\rm 2g}^1$ and $A_{\rm 1g}$ Raman frequencies due to changes in the $c$-axis interactions. Table \ref{table:QualitativeShifts} shows that in Ni-doped MoS$_2$ $A_{\rm 1g}$ has small redshifts for intercalation, like Li-intercalated MoS$_2$, but $A_{\rm 1g}$ has a blueshift for S substitution. Dopants have generally been found to shift mostly the $A_{\rm 1g}$ frequency,\cite{Iqbal} including the measurement of a shift around $-1$ cm$^{-1}$ for $E^1_{\rm 2g}$ and +3 cm$^{-1}$ for $A_{\rm 1g}$ in Ni-doped MoS$_2$;\cite{Kondekar} however, in our calculation we find Ni induces also strong shifts in the $E_{\rm 2g}^1$ peak and the IR-active peaks. The different behavior observed in the experiment may be due to different (and unclear) Ni concentration and the effects of a limited number of layers, in-plane crystallite size, and/or strain.\cite{Kondekar}

Peaks unique to the doped spectra were classified as either ``new'' or ``activations.'' New peaks are those that do not have counterparts in the pristine 3$\times$3 $q$-grid vibrational density of states (VDOS) (Fig. S4) and have mode characters that cannot be easily described by pristine mode eigenvectors: they are mostly related to motions of the Ni atom and its neighbors. Such new peaks were measured in Li-intercalated MoS$_2$ at 205, 1370, and 1600 cm$^{-1}$.\cite{Sekine} Activations are modes that have a counterpart in the pristine case whose IR and/or Raman activity is forbidden by symmetry. Ni-doping breaks symmetries and mixes modes, which can induce IR and/or Raman activity. Raman intensities were particularly large in the case of Mo substitution owing to its metallic character, and so they are scaled down in Fig. \ref{fig:VibSpec}; we expect this intensity enhancement compared to pristine can serve as a feature to identify this kind of doping. There is Raman and IR activity present in all doped structures at low frequencies (below 350 cm$^{-1}$), but it is most prominent in Mo-substituted and o-intercalation structures.

\begin{figure}
	\includegraphics[width=400px]{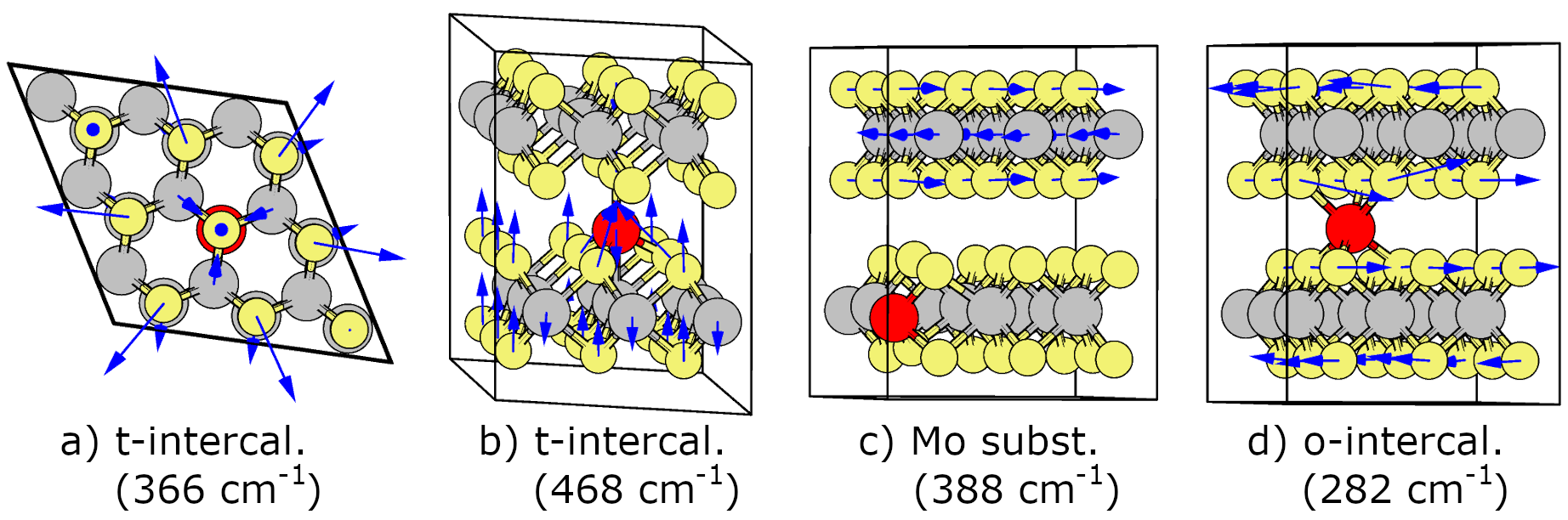}
	\caption{Example vibrational modes of Ni-doped MoS$_2$, showing four typical patterns: a) a new peak for t-intercalation consisting of an in-plane breathing mode around Ni, in only one layer; b) t-intercalation breaks symmetry between the layers, mixing the $A_{\rm 2u}$ and $B^1_{\rm 2g}$ modes with Ni-S stretches; c) Mo substitution breaks symmetry between layers, and mixes $E^1_{\rm 2g}$ and $E_{\rm 1u}$ modes, but does not involve Ni; d) o-intercalation only slightly breaks symmetry, and leaves $E_{\rm 2u}$ modes almost unaffected.}
	\label{fig:VibMode}
\end{figure}

The modes of the doped systems show several interesting patterns in comparison to the pristine modes (Fig. \ref{fig:VibMode}). Many new modes are breathing modes of Mo and S around Ni (Fig. \ref{fig:VibMode}a), or involve Ni-S stretches (Fig. \ref{fig:VibMode}b), or feature new combinations of S motions, as in S substitution or Mo substitution. The activation peaks are strongly influenced by symmetry breaking between the two layers. In Mo and S substitution, only one layer has a dopant, strongly breaking symmetry; pristine modes are mixed to give modes localized on only layer (Fig. \ref{fig:VibMode}c). In t-intercalation, the layer symmetry is broken because Ni has 3 bonds to one layer and 1 to the other, and so the activations involve Mo/S motion in only one layer. For o-intercalation, a perfectly symmetrical structure is possible, but in fact Ni moves slightly closer to one layer, making only a small symmetry breaking: most activated modes (Fig. \ref{fig:VibMode}d) are still on both layers, look identical to pristine, and are shifted typically less than 10 cm$^{-1}$ from the pristine frequencies. Most of the new peaks also show these same layer localization patterns.

Kondekar \textit{et al.}\cite{Kondekar} find a Raman peak related to Ni-doping at 290 cm$^{-1}$.
This peak seems a sign of Mo-substitution, corresponding to our calculated new peak at 300 cm$^{-1}$, which is related to S vibration in the doped layer, out-of-phase between the two S planes. This peak appears for both 2$\times$2 and 3$\times$3 supercells and should be detectable given the high Raman intensity. A less likely possibility (due to lower intensity, and unfavorable energetics) is the o-intercalation, which has a peak at 287 cm$^{-1}$ related to the pristine $E_{\rm 1g}$ (low Raman activity) and $E_{\rm 2u}$ (Raman-inactive) modes. With higher-resolution Raman spectroscopy, we expect that additional peaks could be discerned and solidify the identification.

A few low-frequency modes are particularly relevant to tribology: the shearing mode ($E_{\rm 2g}^2$ in pristine), relating to frictional sliding, and the layer breathing mode ($B_{\rm 2g}^2$ in pristine), relating to layer dissociation and wear. Our calculated values for the doped structures are listed in Table \ref{table:SlidingShearingFreqs}. The increase in the layer breathing frequency in intercalated structures correlates with the stiffening of $C_{33}$ in those structures. Substituted breathing mode frequencies are largely unaffected. The doubly degenerate shearing mode was split in the doped structures, and reduced in frequency for low doping concentration, particularly in S substitution and o-intercalation. At high doping concentrations, both shearing and breathing frequencies are increased.

Raman intensities are related to EPC and have been suggested as a probe of EPC in 2D materials.\cite{LinEPC} In our systems, only Mo substitution shows substantial changes in intensities, whereas other sites stay within 10\% of the pristine peak intensity. If the argument by Lin \textit{et al.}\cite{LinEPC} applies, Mo substitution's large Raman intensity would mean increased EPC and thus increased friction. This means that the observed reduction in friction\cite{Stupp, Vellore} of Ni-doped MoS$_2$ is either not attributable to changes in EPC or samples were not dominated by substitution of the Mo site.

\begin{figure}
	\includegraphics[width=450px]{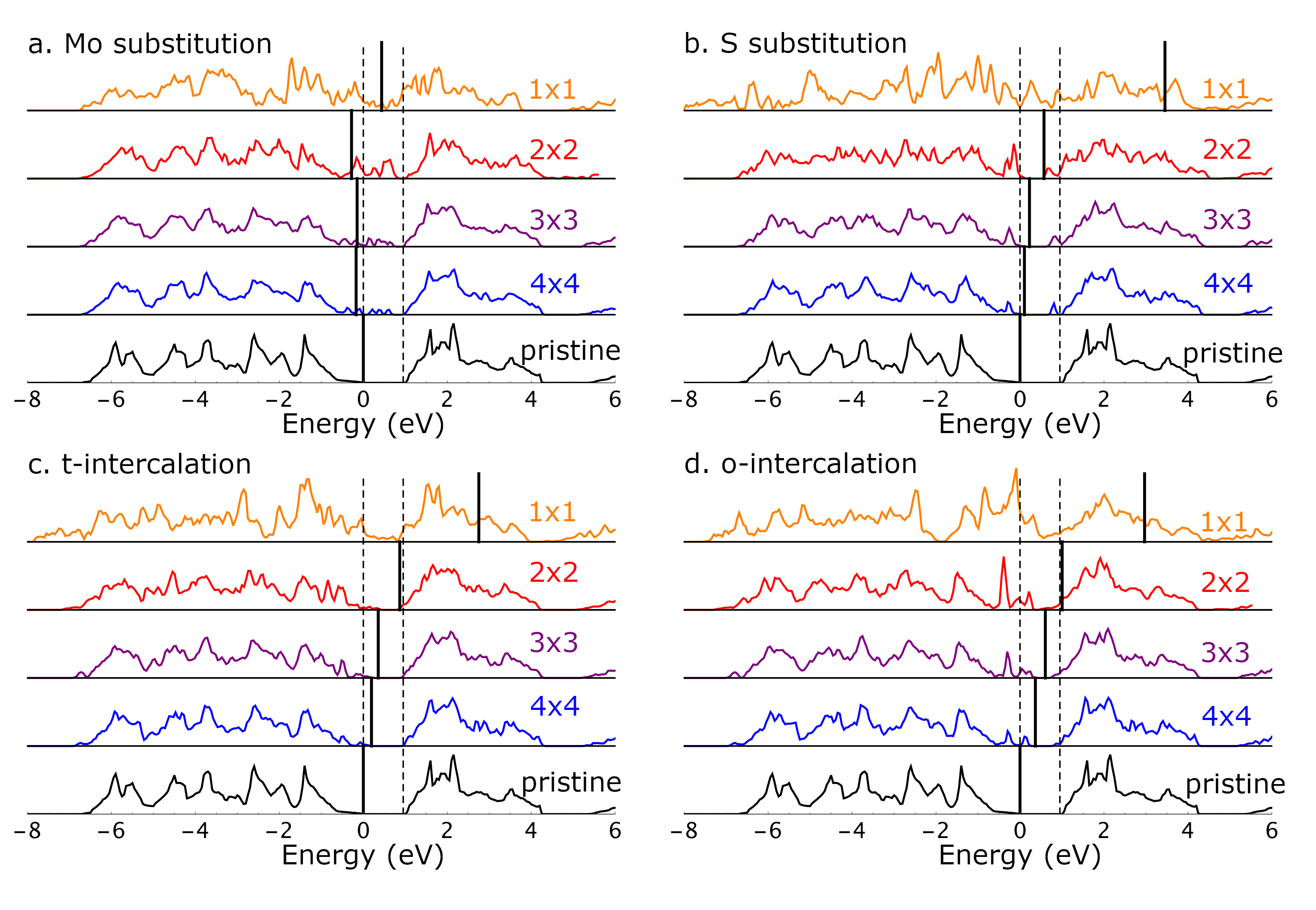}
	\caption{Electronic density of states (DOS) for doped structures (color) and pristine DOS (black). The CBM and VBM of pristine MoS$_2$ are shown as vertical dashed lines. Doped DOS were aligned with the low-lying Mo 4$s$ state (36 eV below VBM) of the pristine plot. Calculated Fermi energies or VBM are shown as solid-black vertical lines.}
	\label{fig:EDOS}
\end{figure}

To assess $n$-type or $p$-type doping and other changes in electronic structure, we calculate the electronic density of states (DOS), shown in Fig. \ref{fig:EDOS}. Low concentrations of S-substituted structures show defect states near the conduction band maximum (CBM) while o-intercalation creates defect states above the valence band maximum (VBM). All doped structures show some degree of bandgap closing. Mo-substituted structures have metallic character, with the Fermi level situated in the new in-gap states meaning it is slightly $p$-type. S substitution shows new states near both band edges, while o-intercalation has new states near the valence band and an extended conduction band tail. The t-intercalation shows the least modification from the pristine DOS, with only a slight narrowing of the gap due to new states in the valence band.

The Fermi level for low doping concentration is near the pristine VBM. Alterations are more dramatic at high concentrations where the Fermi level is pushed well within the pristine conduction band in all cases except Mo substitution. This can be understood as $n$-type doping at high concentration. Naturally occurring MoS$_2$ is typically $n$-type due to Re impurities.\cite{Komsa} Provided nickel sulfides do not form,\cite{Kondekar} high Mo-substitution doping concentrations could create $p$-doped MoS$_2$, which has been challenging to achieve.\cite{Vandalon}

In general, our calculations indicate an increased conductivity of Ni-doped MoS$_2$, either through in-gap states, a narrowed bandgap, or $n$- or $p$-type electronic doping, depending on the doping site. These results are consistent with enhanced catalytic activity of Ni-doped MoS$_2$, and early calculations attributing enhancement in Mo-substituted MoS$_2$ to electron donation from Ni to Mo.\cite{Gomez}

\section{Conclusion}
Through \textit{ab initio} methods, we have computed the structure, stability, and vibrational spectra of bulk Ni-doped 2H-MoS$_2$, which has not been previously explored. We found four metastable dopant sites: Mo substitution, S substitution, octahedral intercalation, and tetrahedral intercalation (always favored energetically over octahedral). Ni in the Mo-substituted structure has only five bonds just as in a basal plane surface,\cite{Hakala} and thus is elastically weaker in-plane. The $C_{55}$ shear-stiffness parameter is strengthened by doping, suggesting this is not the explanation for the reduced friction observed\cite{Stupp} in Ni-doped MoS$_2$. Ni-S bonding is comparable to pristine Mo-S bonding, and the fact that Ni-Mo bonds are comparable to Mo-S bonds in S substitution are particularly surprising---this suggests that Ni, like Co,\cite{Park} can fill existing S vacancies as a way to synthesize S-substituted MoS$_2$.

We find that all doped structures are above the convex hull, but t-intercalation is very close to it. It therefore can be considered thermodynamically stable, with essentially zero formation energy from Ni and MoS$_2$, and is not likely to phase-segregate, unlike the other doped structures. The other sites' energies above hull are lower than that of the known Chevrel phase NiMo$_3$S$_4$, suggesting these are synthesizable as well. Phase diagrams show how the chemical potentials of Mo and S can be tuned in synthesis to favor Mo substitution, S substitution, or t-intercalation, though only t-intercalation can co-exist with pristine MoS$_2$. Intercalated Ni forms strong covalent bonds between layers, unlike intercalated Li. This increases elastic parameters out of plane, and results in little change to the $c$-parameter, contrary to what is sometimes assumed. Strong interlayer bonding poses a possible explanation for the observed resistance to wear.\cite{Stupp, Vellore}

$n$-type doping is prominent at high Ni concentration. Mo-substituted structures become metallic due to in-gap states, which leads to a large enhancement of Raman intensities. New peaks in the Raman spectra appear due to vibrations of Ni and its neighbors, while activations and shifts in existing modes are caused by breaking symmetry and altering bond strengths. While generally doping has been found primarily to affect the $A_{\rm 1g}$ mode, we find shifts of all four IR- and Raman-active peaks. The distinctive effects on the Raman and IR spectra can be used experimentally to identify the doping sites in Ni-doped samples, particularly with high-resolution, tip-enhanced, or resonant Raman spectroscopy. We propose a new paradigm for identification of MoS$_2$ dopant locations by vibrational spectroscopy, as these structures have remained unclear experimentally.

The structures found in this work are being used to probe macroscopic tribological properties related to wear and friction through parametrization of ReaxFF force fields\cite{Ostadhossein} for classical molecular dynamics and calculations of sliding potentials.\cite{Liang} We are additionally extending this work to study Ni-doping in 3R bulk and 1H and 1T monolayer polytypes of MoS$_2$\cite{Karkee} to determine the impact of the phase on doping effects.

\section{Methods}
\subsection{DFT Calculations}
We use plane-wave density functional theory (DFT) and density functional perturbation theory\cite{Baroni} (DFPT) implemented in the code \texttt{Q\textsc{uantum}} \texttt{ESPRESSO}\cite{QE, QE2009} version 6.4. Calculations were performed using the Perdew-Burke-Ernzerhof\cite{PBE} (PBE) generalized gradient approximation and Perdew-Wang\cite{PW} local density approximation (LDA) exchange-correlation functionals with optimized norm-conserving Vanderbilt pseudopotentials\cite{Hamann} parametrized by Schlipf and Gygi\cite{SG15} (for PBE) and by PseudoDojo\cite{PseudoDojo} (for LDA) obtained from their respective websites.\cite{PseudoDojo, SG15}
Raman calculations were done only for LDA since PBE is not compatible with Raman intensity calculations in this code.\cite{Lazzeri}

Relaxations and electronic structure of the primitive, 6-atom MoS$_2$ cells used a $k$-point grid of $6\times6\times4$ with a half-shifted $k$-grid. Increasing the supercell size to reach lower Ni concentrations allows decreasing $k$-points per axis in the in-plane $a$- and $b$-directions inversely proportional to the supercell size: $4\times4\times4$, $3\times3\times4$, and $2\times2\times4$ $k$-points per axis were used for $2\times2\times1$, $3\times3\times1$, and $4\times4\times1$ supercells, respectively. Atomic coordinates were relaxed using force thresholds of 10$^{-4}$ Ry/Bohr and the stresses were relaxed to 0.1 kbar. Calculations were spin-unpolarized, except for ferromagnetic bulk Ni. Spin-polarization for other cases was found to affect the total energy by less than 0.001 meV per atom, thus having no significance for the properties considered here. The change is only non-negligible in bulk Ni, where the energy difference is 0.2 eV/atom.

A wavefunction cutoff of 60 Ry was used for PBE and 80 Ry for LDA. The self-consistency thresholds were set to 10$^{-18}$ for the ground state and 10$^{-15}$ for phonons. Strict thresholds were required for accurate calculations of modes with low frequencies. Low frequencies of vibrational calculations of 3$\times$3 t-intercalation and Mo substitution were initially calculated to be imaginary and required special care to properly converge: $k$-points were increased to $4\times4\times4$ and the phonon self-consistent threshold was lowered to 10$^{-16}$. DOS calculations were carried out on fine $k$-grid meshes of $12\times12\times12$, $10\times10\times10$, $6\times6\times6$, and $6\times6\times6$ for $1\times1\times1$, $2\times2\times1$, $3\times3\times1$, and $4\times4\times1$ supercell sizes, respectively.

\subsection{Elasticity}

Elastic parameters were calculated using the stress-strain relationship (using Voigt notation) $\sigma_{i} = C_{ij} \epsilon_{j}$. We applied uniaxial strains in the 1- and 3-directions and 5-direction shear strains to calculate $C_{11}$, $C_{33}$, and $C_{55}$. Strains were applied from -0.01 to 0.01 in intervals of 0.002, with each structure's atoms relaxed while holding the lattice vectors constant. $C_{ij}$ was determined by linear regression on the stress \textit{vs}. strain. Elastic parameters for pristine structures using different functionals are listed in Table \ref{table:Fnl}, demonstrating good agreement for LDA and PBE+GD2 with experiment.

Strain directions 3 and 5 were chosen for their relevance to sliding and wear. Shearing strains in the 4- and 5-directions are the motions involved in basal plane sliding. Uniaxial 3-strain is involved in layer separation, leading to wear. By symmetry, $C_{11} = C_{22}$ and $C_{44} = C_{55}$ in pristine MoS$_2$, and these relations still approximately hold in our doped structures.

\subsection{Formation Energy}
Formation energies, $E_{{\rm form}}$, can be used to gauge the relative stability of a structure versus a reference,\cite{Kim, Ivanovskaya} and are defined as:
\begin{equation} \label{eqn:eform}
E_{{\rm form}} = E_{{\rm mixed}} - \sum_{i} N_i \mu_{i}
\end{equation}
$E_{\rm mixed}$ is the energy of the material of interest, $\mu_{i}$ is the chemical potential of the bulk element $i$, and $N_i$ is the number of atoms of element $i$ in the mixed system. $\mu_i$ must be less than the bulk element's energy per atom for the formation of the material of interest to be thermodynamically favored compared to its bulk elements, $E_{i}$. For this reason, it is useful to write Eqn. \ref{eqn:eform} instead using $\Delta \mu \equiv \mu_{i} - E_{i}$, so
\begin{equation} \label{eqn:eform2}
E_{{\rm form}} = E_{{\rm mixed}} - \sum_{i} N_i (\Delta \mu_{i} + E_{i})
\end{equation}

Formation energies were all referenced against the most stable bulk elemental compounds: Fm$\bar{3}$m (fcc) Ni, Im$\bar{3}$m (bcc) Mo, and P2$_1$ S. S is a particularly difficult element to handle, given its natural state is 8-membered rings. This S structure was chosen as it has been used in other literature\cite{Rasmussen, Dolui} and had the lowest calculated energy per atom among an isolated S atom, S$_2$ molecule, an isolated 8-membered ring, and an arrangement of 4 8-membered rings per cell. The reference elemental structures are metals in the case of Ni and Mo while the structures of interest are mostly semiconducting. This leads to the required use of smearing on the electronic states of some of the structures. Furthermore, some structures have significant Van der Waals interactions while others do not. These differences motivate LDA for direct comparison of formation energies rather than PBE or PBE + GD2. Lattice parameters, Materials Project IDs, $k$-grids, and space group symmetries for all of the mentioned structures can be found in Table S1. Formation energies are shown in Table S4.

Using Eqn. \ref{eqn:eform2}, we can compare the formation energies of doped structures to construct zones of chemical potentials that favor the formation of one structure over another at a given supercell size as in Fig. \ref{fig:chemPot}. To find these regions, we need to solve for when $E_{{\rm form}} (\Delta \mu_{\rm Ni}, \Delta \mu_{\rm Mo},\Delta \mu_{\rm S})$ of one structure is smaller than another. Since each computation only contains one Ni atom per supercell, we can remove the dependence of $\Delta \mu_{\rm Ni}$, leaving us with zones as a function of only $\Delta \mu_{\rm Mo}$ and $\Delta \mu_{\rm S}$.

We represent phase stability using the concept of the convex hull\cite{Ong} which we compute using Mathematica.\cite{Mathematica} Considering the closed system ($\Delta \mu = 0$) at zero temperature and pressure, we can use the formation energies to construct the convex hull as the smallest convex set containing the points ($x_{\rm Ni}$, $x_{\rm Mo}$, $x_{\rm S}$, $E_{\rm form}/N_{\rm atoms}$), where $x_{i} = N_{i}/(N_{\rm atoms})$.\cite{Ong} Since $\Sigma_i x_i=1$, we can represent the three $x_i$ dimensions as three legs of the two-dimensional triangle, as in Fig. \ref{fig:ConvexHull}. The $E_{\rm form}/N_{\rm atoms}$ axis is out of the page, but we can further simplify this representation by owing only the lines connecting points on the convex hull. Stable structures will be on the convex hull's boundary. Any structure whose computed energy is above the convex hull is favored to phase-segregate to the structures at the nearest nodes. The energies above hull are listed in Table \ref{table:EnAboveHull} for LDA and PBE+GD2 in Table S3. One-dimensional convex hull diagrams with reference structures that are not bulk elemental phases are shown in Fig. S2.

\begin{suppinfo}
Phase diagrams and convex hull diagrams for LDA, PBE, and PBE+GD2;
convex hull diagrams with non-elemental references;
formation energies with respect to bulk elemental structures and filling pre-existing vacancies;
Materials Project ID, $k$-grid, and cell parameters for the structures besides Ni-doped MoS$_2$ featured in this work;
and vibrational density of states and vibrational mode characterizations (PDF).
XSF file format structures of Ni-doped MoS$_2$ structures relaxed with PBE+GD2 (ZIP).
AXSF file format structures of Ni-doped MoS$_2$ vibrational eigenmodes, calculated with LDA (ZIP).
\end{suppinfo}

\begin{acknowledgement}
We acknowledge useful discussions with Mehmet Z. Baykara and Ashlie Martini. We acknowledge support from UC Merced start-up funds and from the Merced nAnomaterials Center for Energy and Sensing (MACES), a NASA-funded research and education center, under award NNX15AQ01. This work used computational resources from the Multi-Environment Computer for Exploration and Discovery (MERCED) cluster at UC Merced, funded by National Science Foundation Grant No. ACI-1429783, and the National Energy Research Scientific Computing Center (NERSC), a U.S. Department of Energy Office of Science User Facility operated under Contract No. DE-AC02-05CH11231.
\end{acknowledgement}


\providecommand{\latin}[1]{#1}
\makeatletter
\providecommand{\doi}
  {\begingroup\let\do\@makeother\dospecials
  \catcode`\{=1 \catcode`\}=2 \doi@aux}
\providecommand{\doi@aux}[1]{\endgroup\texttt{#1}}
\makeatother
\providecommand*\mcitethebibliography{\thebibliography}
\csname @ifundefined\endcsname{endmcitethebibliography}
  {\let\endmcitethebibliography\endthebibliography}{}

\section*{TOC Graphic}
\label{For Table of Contents Only}
\centering
\includegraphics[]{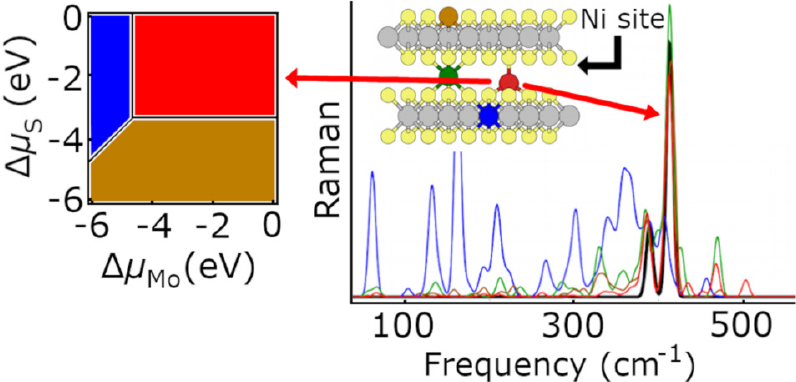}

\end{document}